# Fabrication of Ag/Tl–Ba–Ca–CuO/CdSe nanostructure by electro-deposition technique

P.M. Shirage, D.D. Shivagan, L.A. Ekal, N.V. Desai, S.B. Mane, S.H. Pawar[*]

Department of Physics, School of Energy Studies, Shivaji University, Kolhapur 416004, India

## Abstract

Junctions between metal, semiconductor and superconductor materials seem to be promising for their application as electronic devices, microelectronic devices and high power transmission. We fabricated Ag/Tl–Ba–Ca–CuO/CdSe nanostructures by using dc electro-deposition technique. The Tl–Ba–Ca–Cu alloyed thin films were deposited onto silver substrate at a constant potential of $-1.25$ V and then oxidized electrochemically at a potential of $+0.7$ V with respect to saturated calomel electrode (SCE) from 1 N KOH solution. Semiconducting CdSe was subsequently deposited onto superconducting system at a constant potential of $-0.6$ V with respect to SCE. Particle size confirmed the presence of nanostructure, of the order of 27 nm. The sizeable change of the electrical conductivity was obtained by irradiating the nanostructure with red He–Ne laser ($\lambda = 632.8$ nm) for different periods. © 2001 Elsevier Science B.V. All rights reserved.

## 1. Introduction

For most of the power related and microelectronic based applications, superconducting thin films/wires with $J_c$ values of the order of $10^5$ A/cm$^2$ are required. The $J_c$ values can be increased by enhancing the carrier concentration and their mobilities. These parameters can be monitored by controlling the preparative conditions of the superconducting thin films/wires. On the other hand, photosensitive materials irradiated with light significantly increase the electrical conductance. Devices with both of these properties can be achieved by forming a junction of photosensitive semiconductor with superconductors. Such type of structures were fabricated with Ag/Bi–Sr–Ca–CuO/CdSe and studied for their electrical properties at different low temperatures. It has been reported that the $T_c$ values can be raised as high as 250 K with such structures [1,2].

For device applications, the Tl-based superconducting oxides present an attractive alternative to BSCCO and YBCO due to number of features including high transition temperature at about 127 K. In view of this, Tl-2223 system has been selected as superconducting system for the fabrication of nanostructure with photosensitive semiconductor. In the present investigation, CdSe semiconductor has been selected as it possess good photosensitive and nanocrystalline properties.

In the present paper, we are reporting the fabrication of Ag/Tl–Ba–Ca–CuO/CdSe multi-layer structure using electro-deposition technique. Fabrication was confirmed by studying X-ray diffraction (XRD), FWHM and SEM. The change in I–V characteristics before and after laser irradiation were studied to observe change in the normal state resistance of this junction.

## 2. Experimental

In the present investigation, the multi-layered structure of Ag/Tl-2223/CdSe was fabricated by dc electro-deposition technique. The electro-deposition allows to deposit all the ionic constituents in the required controlled stoichiometric fashion. The Tl–Ba–Ca–Cu alloy films were deposited by using electro-chemical technique and are reported elsewhere [3]. The alloyed films were oxidized further in electro-chemical cell, with alloyed Tl–Ba–Ca–Cu film as working electrode and 1 N KOH solution as an electrolyte. The electrochemical potential acts as the driving force to intercalate oxygen species into alloyed layers. A layer of CdSe semiconductor was then electro-chemically deposited on Tl–Ba–Ca–CuO electrode using a aqueous solution of $CdSO_4$ (50 mM) and $SeO_2$ (10 mM). The potential of $-0.6$ V, with respect to saturated calomel electrode (SCE) in electro-chemical cell, was maintained and CdSe films were deposited for the period of 20 min.

XRD data of multi-layered structure was obtained on microcomputer controlled Philips PW-3710 diffractometer using Cu $K\alpha$ radiation. FWHM for this film was studied and the particle size was calculated by using Scherrer's formula. The microstructure was observed using a close circuit television attachment to Metzer optical microscope (450×) under reflection mode and by using SEM probe microscopy with CAMECA Model SU 30. The I–V characteristics are a key parameter to determine the conductivity of the surfaces and interfaces. In order to study the electrical properties of this structure, the contacts were made from the deposit surface and Ag substrate, by air drying silver paint. The I–V characteristics were studied in dark and under photoexcitation.

## 3. Results and discussion

The Tl–Ba–Ca–Cu alloy was successfully deposited at potential of $-1.25$ V with respect to SCE. The thickness of the film deposited for 15 min duration was found to be of the order of 2 μm. The optimized potential of electro-chemical oxidation was found to be +0.7 V with respect to SCE and the period of oxidation of Tl–Ba–Ca–Cu alloy is 28 min, at which sample showed superconductivity below transition temperature of 114 K [4].

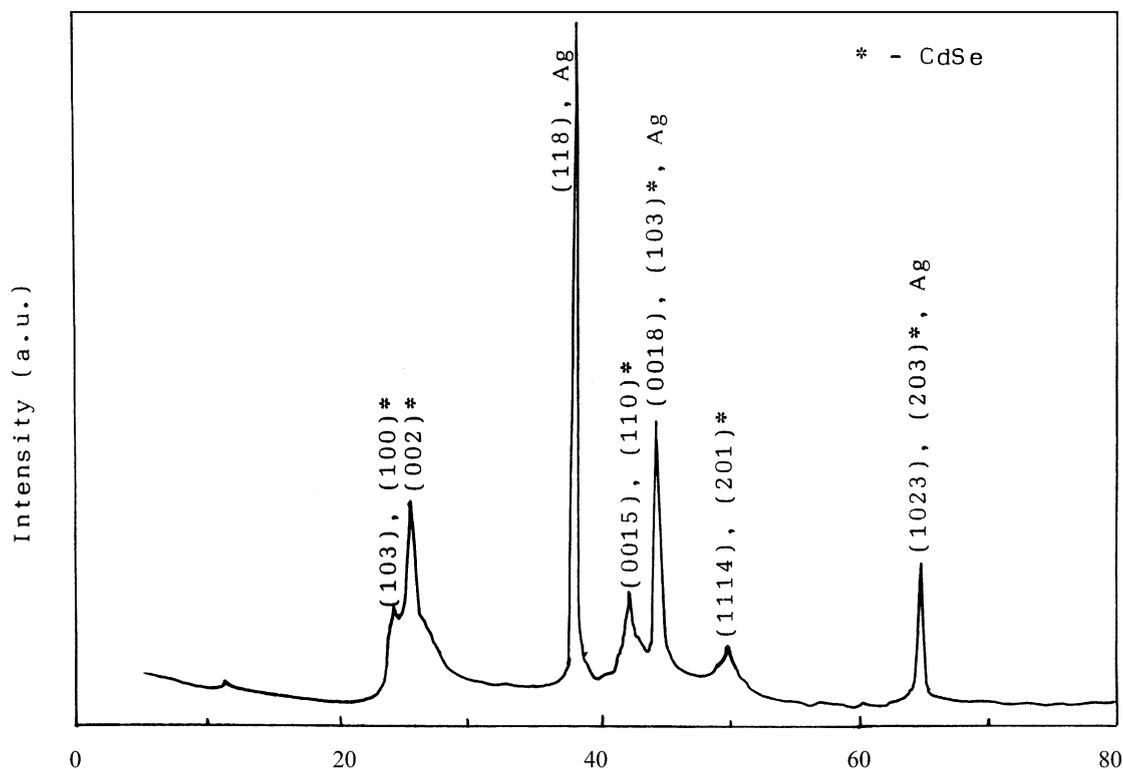

Fig. 1. XRD pattern of Ag/Tl–Ba–Ca–CuO/CdSe multi-layered nanostructure.

The thickness of CdSe deposited onto the superconductor, Tl-2223 system for 20 min is 1.5 μm. The XRD pattern of this multi-layered nanostructure is shown in Fig. 1. The pattern consists of tetragonal superconducting Tl-2223 system and hexagonal semiconducting CdSe with standard lattice parameters, this confirms the fabrication of multi-layer structure. The particle size was calculated by using Scherrer's formula applied to FWHM and was found to be 27 nm. The microstructure of the alloyed, oxidized superconducting system and Ag/Tl-2223/CdSe nanostructure was observed by Metzer optical microscope and it was found that films were uniform and well covered. The multi-layered structure was then characterized by SEM and typical one is shown in Fig. 2.

The photo-induced changes in *I–V* characteristics of Ag/Tl–Ba–Ca–CuO/CdSe multi-layered nanostructure were studied, at room temperature, in dark and in presence of laser irradiation. The red He–Ne laser of 2 mW power with $\lambda = 632.8$ nm and the photon energy greater than the band gap of CdSe and superconducting system was used. This power is sufficiently high and could enhance the electronic excitations by producing electron–hole pairs, along with micro-level heating of the sample at the irradiated area.

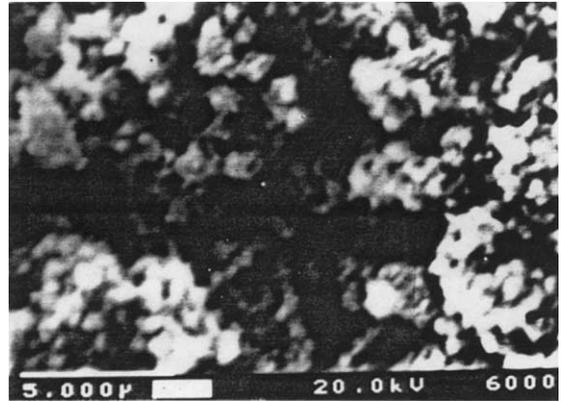

Fig. 2. Typical SEM photograph of Ag/Tl–Ba–Ca–CuO/CdSe structure.

The recorded data of *I–V* measurements are shown in Fig. 3. The plots are observed to be linear. It is seen that, slope of *I–V* curve is found to increase with laser irradiation. Further, the increase in the period of irradiation increases the slope of the *I–V* plot and became steady after 3 h. This is attributed to the resultant decrease in the normal state resistance, which might have been caused due to increase in the carrier concentration. The laser heating effect seems to be very minor as there is a dissipation of heat through silver substrate used in multi-layered structure. The increase in conductivity shows that laser irradiation assists in enhancing the carrier concentrations and their mobilities. These carriers drifted according to the applied potential to the junction and hence resulting in increase in conductivity. This will help to increase the superconducting parameters such as $T_c$ and $J_c$ values. Low temperature studies of this work is in progress.

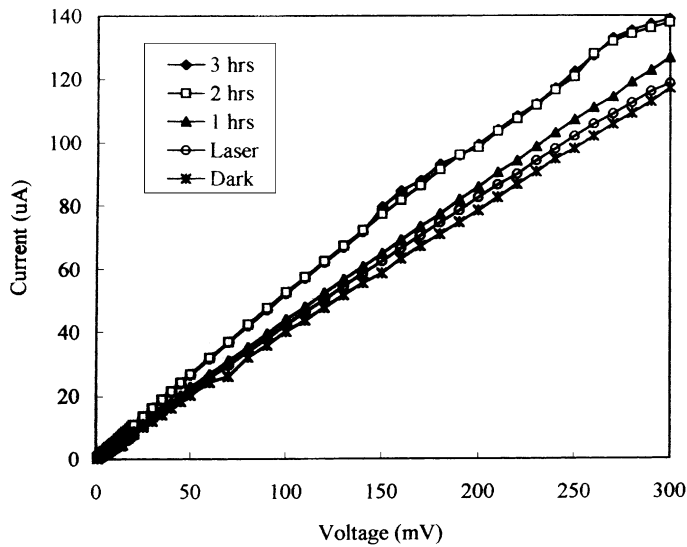

Fig. 3. Plot of *I–V* characteristics of Ag/Tl–Ba–Ca–CuO/CdSe multi-layered nanostructure in dark and with laser irradiation.

In conclusion, the multi-layered nanostructures of metal–superconductor–semiconductor has been successfully fabricated by electro-deposition technique. The increase in slope ($dI/dV$) after laser irradiation, shows the decrease in the normal state resistance of the junction. Thus, superconductor–semiconductor multi-layered nanostructures could offer superior properties than conventional materials.

## Acknowledgements

Author wish to thank Dr. A.V. Narlikar for his constant encouragement and UGC, New Delhi for financial support.

## References

[1] T.S. Desai, Ph.D. Thesis, Shivaji University, Kolhapur, 1995.
[2] S.H. Pawar, T.S. Desai, L.A. Ekal, D.D. Shivagan, P.M. Shirage, Int. J. Inorg. Mater., in press.
[3] L.A. Ekal, N.V. Desai, S.H. Pawar, Bull. Mater. Sci. 22 (4) (1999) 775–778.
[4] L.A. Ekal, P.M. Shirage, D.D. Shivagan, N.V. Desai, S.B. Kulkarni, S.H. Pawar, Int. J. Ceram., submitted for publication.